\documentclass[fleqn]{2020SCGE}
\usepackage{epstopdf}
\usepackage{cite}
\usepackage{graphics}
\begin{document}

\ensubject{subject}


\ArticleType{Article}
\SpecialTopic{SPECIAL TOPIC: The Chinese H$\alpha$ Solar Explorer -- CHASE Mission}
\Year{2022}
\Month{March}
\Vol{65}
\No{??}
\DOI{10.1007/s11433-022-1900-5}
\ArtNo{289603}
\ReceiveDate{March 4, 2022}
\AcceptDate{March 28, 2022}

\title{Calibration procedures for the CHASE/HIS science data}{Calibration procedures for the CHASE/HIS science data}

\author[1,2]{Ye Qiu}{}%
\author[1,2]{ShiHao Rao}{}
\author[1,2]{Chuan Li}{{lic@nju.edu.cn}}%
\author[1,2]{Cheng Fang}{}
\author[1,2]{MingDe Ding}{{dmd@nju.edu.cn}}
\author[1,2]{Zhen Li}{}
\author[1,2]{\\YiWei Ni}{}
\author[1,2]{WenBo Wang}{}
\author[1,2]{Jie Hong}{}
\author[1,2]{Qi Hao}{}
\author[1,2]{Yu Dai}{}
\author[1,2]{PengFei Chen}{}
\author[1,2]{XiaoSheng Wan}{}
\author[3]{\\Zhi Xu}{}
\author[4]{Wei You}{}
\author[4]{Yuan Yuan}{}
\author[5]{HongJiang Tao}{}
\author[5]{XianSheng Li}{}
\author[5]{YuKun He}{}
\author[5]{Qiang Liu}{}

\AuthorMark{Y. Qiu, S. H. Rao, C. Li, C. Fang, M. D. Ding, Z. Li, et al.}

\AuthorCitation{Y. Qiu, S. H. Rao, C. Li, C. Fang, M. D. Ding, Z. Li, et al.}

\address[1]{School of Astronomy and Space Science, Nanjing University, Nanjing 210023, China}
\address[2]{Key Laboratory for Modern Astronomy and Astrophysics (Nanjing University), Ministry of Education, Nanjing 210023, China}
\address[3]{Yunnan Observatories, Chinese Academy of Sciences, Kunming 650216, China}
\address[4]{Shanghai Institute of Satellite Engineering, Shanghai 201109, China}
\address[5]{Changchun Institute of Optics, Fine Mechanics and Physics, University of Chinese Academy of Sciences, Changchun 130033, China}

\abstract{The H$\alpha$ line is an important optical line in solar observations containing the information from the photosphere to the chromosphere. To study the mechanisms of solar eruptions and the plasma dynamics in the lower atmosphere, the Chinese H$\alpha$ Solar Explorer (CHASE) was launched into a Sun-synchronous orbit on October 14, 2021. The scientific payload of the CHASE satellite is the H$\alpha$ Imaging Spectrograph (HIS). The CHASE/HIS acquires, for the first time, seeing-free H$\alpha$ spectroscopic observations with high spectral and temporal resolutions. It consists of two observational modes. The raster scanning mode provides full-Sun or region-of-interest spectra at H$\alpha$ (6559.7 -- 6565.9 \AA) and Fe I (6567.8 -- 6570.6 \AA) wavebands. The continuum imaging mode obtains full-Sun photospheric images at around 6689 \AA. In this paper, we present detailed calibration procedures for the CHASE/HIS science data, including the dark-field and flat-field correction, slit image curvature correction, wavelength and intensity calibration, and coordinate transformation. The higher-level data products can be directly used for scientific research.}

\keywords{Space-based telescope, Solar physics, Chromosphere, Photosphere}
\PACS{95.55.Fw, 96.60.--j, 96.60.Na, 96.60.Mz}

\maketitle


\begin{multicols}{2}
\section{Introduction}\label{introduction}

Since the very first observations of solar H$\alpha$ telescopes \cite{hale}, the H$\alpha$ line has always been an important line used in the spectral and imaging solar observations. In quiet Sun, the H$\alpha$ line is among the strongest and broadest optical lines. In the regions of solar activities like solar flares, the H$\alpha$ line can become an emission line with its strength related to the heating extent of the plasma where the line originates. The images at the H$\alpha$ center reflect the chromospheric characteristics and the quiescent filaments, while the far wings carry information of the photosphere, as well as the erupting filaments \cite{chen1}. The H$\alpha$ line asymmetries, mostly appearing in solar flares, reflect the plasma dynamics in the chromosphere. Therefore, the variations of the line profile and the deduced Doppler velocities provide information of the local plasma temperature and nonthermal motions \cite{ichimoto,Berlicki,Madjarska,hong}.

\begin{figure*}[t]
	\centering
	\includegraphics[scale=1.0]{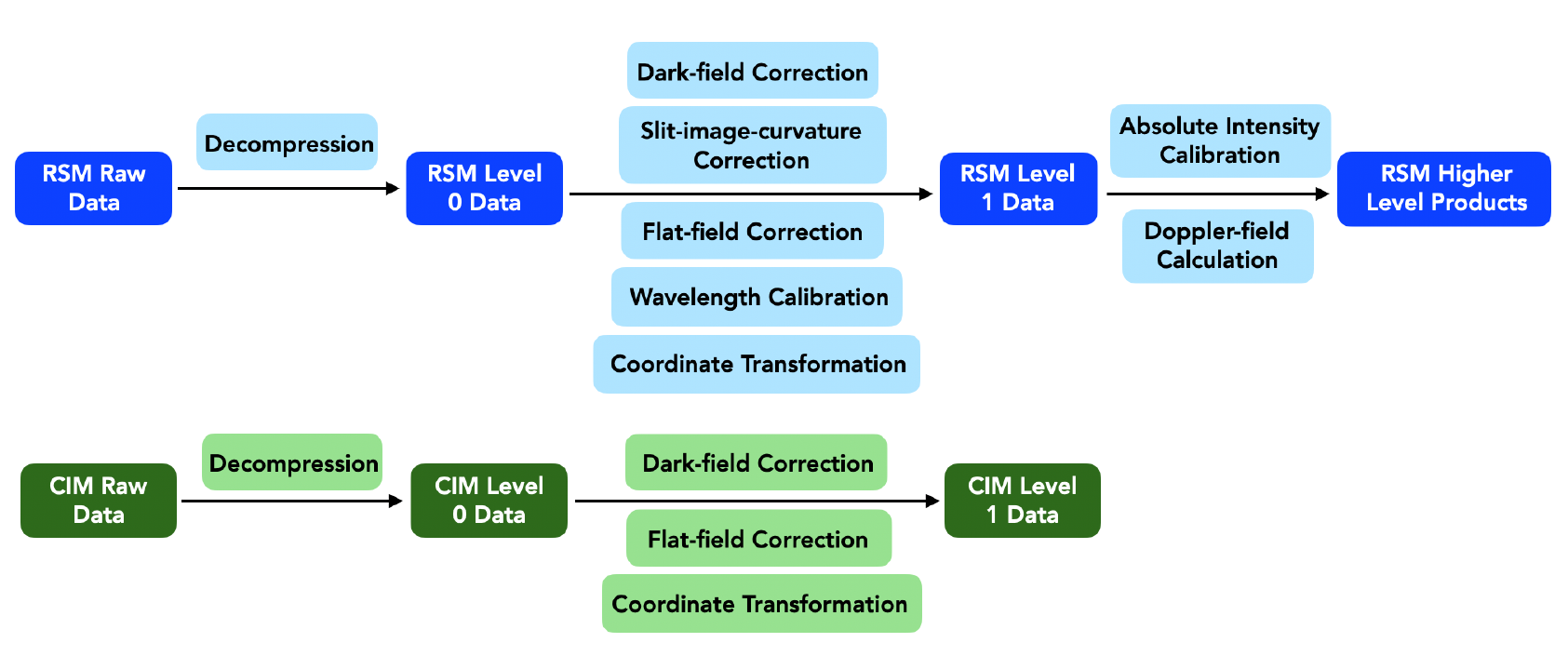}
	\caption{The data calibration flow diagram for the two observational modes of the CHASE/HIS.} 
	\label{fig:flow}
\end{figure*}

It is well known that the ground-based solar telescopes, such as the Optical and Near-infrared Solar Eruption Tracer (ONSET) \cite{fang1}, the New Vacuum Solar Telescope (NVST) \cite{liu}, the Kanzelh$\rm \ddot{o}$he Observatory (KSO) \cite{potzi}, and the under-construction 2.5m Wide-field and High-resolution Solar Telescope (WeHoST) \cite{fang2}, suffer from the seeing and weather effects arising from the Earth's atmosphere, and cannot provide all-day observations. Thus it is important to conduct solar observations in space, for instance, the Chinese space missions such as the FY-3E satellite \cite{zhangp}, the Chinese H$\alpha$ Solar Explorer (CHASE) \cite{li1,li2}, and the to-be-launched Advanced Space-based Solar Observatory (ASO-S) \cite{gan}.

The CHASE mission aims to test an ultra-high precision and stability satellite platform \cite{zhang1}, and to perform solar H$\alpha$ spectroscopic observations that are important for exploring the plasma dynamics in the solar lower atmosphere. Observations of the full Sun are also useful in the Sun-as-a-star studies that provide implications to stellar physics \cite{li2}. The scientific payload onboard the CHASE satellite is the H$\alpha$ Imaging Spectrograph (HIS) \cite{han}. Since its launch on October 14, 2021 and the first-light imaging on October 24, 2021, the in-orbit performance of CHASE/HIS has been excellent. This paper presents detailed calibration procedures from the raw data to the higher-level products, and aims to help users better understand the science data of the CHASE mission.

\section{CHASE data overview}\label{sect2}

The CHASE/HIS has two observational modes: the raster scanning mode (RSM) and the continuum imaging mode (CIM). The RSM acquires the spectra of full-Sun or region-of-interest in two wavebands of H$\alpha$ (6559.7 -- 6565.9 \AA) and Fe I (6567.8 -- 6570.6 \AA) with high spectral and temporal resolutions. The broad range of wavelength helps capture the phenomena occurring from the photosphere to the chromosphere, and the high spectral resolution allows obtaining fine structures in the line profile and deriving the Doppler velocities of the solar lower atmosphere precisely. The temporal resolutions of the full-Sun scanning, region-of-interest scanning and sit-stare spectroscopy are 1 minute, 30 -- 60 seconds and \textless10 ms, respectively. The CIM acquires photospheric images at the continuum around 6689 $\AA$ with a full width at half maximum (FWHM) of 13.4 $\AA$ and a temporal resolution of 1 second. The CIM is designed for observing solar activities in the photosphere and testing the stability of the CHASE satellite platform.

The raw data of the CHASE/HIS are JPEG2000 compressed in orbit. They are transmitted to three ground stations (Miyun, Kashi, and Sanya) located in China, and transferred to the Solar Science Data Center of Nanjing University (SSDC-NJU) through dedicated internet access. The Level 0 data are decompressed from the raw data, and then calibrated to the Level 1 data that can be directly used for scientific purposes. The higher-level products, such as the absolute-intensity spectra and the Dopplergrams, can be further derived based on the RSM Level 1 data. Figure \ref{fig:flow} shows the calibration flows from the raw data to the Level 1 data or the higher-level products for the two observational modes. The following Section \ref{sect3} and Section \ref{sect4} describe the detailed pipelines for the RSM and CIM data, respectively.

\section{Processing pipeline for RSM data}\label{sect3}

The RSM Level 0 spectra are archived as 4608 $\times$ 376 arrays, where the number 4608 refers to the pixels along the slit and 376 refers to the pixels of wavelength. A full-Sun scanning produces $\sim$4625 spectra totally and generates $\sim$14.9 GB data. The Level 0 spectra cannot be directly used for scientific research. Processing RSM Level 0 data to the Level 1 science data or higher-level products involves several steps. The following subsections describe in detail the processing pipeline for RSM data.

\subsection{Dark field}

The dark field is the signal output of the detector without external light source. It is mainly composed of bias (also named digital offset), readout noise and dark current. The first two are the inherent properties of the detector, while the dark current arises from the thermal motion of electrons.

\begin{figure}[H]
	\centering
	\includegraphics[scale=0.3]{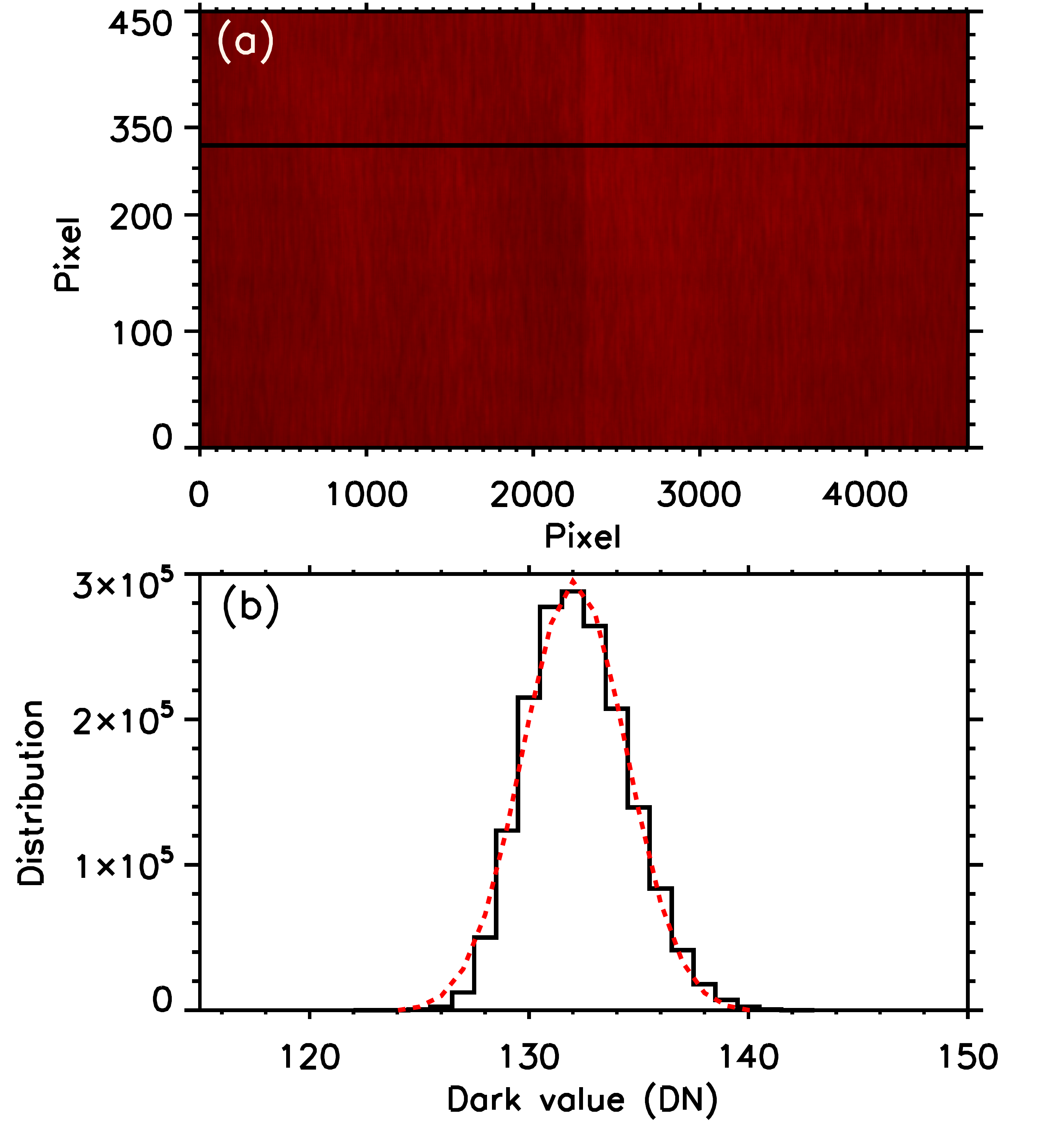}
	\caption{(a) The dark field of RSM obtained on October 24, 2021. (b) The normal distribution of dark field values.} 
	\label{fig:dark}
\end{figure}

We have obtained the dark field of RSM with the satellite pointing to a cold space away from the Sun. The HIS filter assembly and the slit-raster system ensure that the light from other stars hardly produces signals to the detector. The dark field, as shown in Figure \ref{fig:dark}(a), is an average of over 1000 images taken on October 24, 2021. The exposure time of the dark-field images was 5 ms, and the working temperature of the CMOS detector was 13.25 $\rm^{\circ}C$. The RSM detector applies a window of 4608 pixels along the slit and 376 pixels along the wavelength. The black line in Figure \ref{fig:dark}(a) separates the two passbands of H$\alpha$ and Fe I. The normal distribution of dark field values is shown in Figure \ref{fig:dark}(b). The mean value is $\sim$132.3 DN that is about 3$\%$ of the maximum DN according to the 12 bit quantization of the CMOS detector.

\begin{figure}[H]
	\centering
	\includegraphics[scale=0.35]{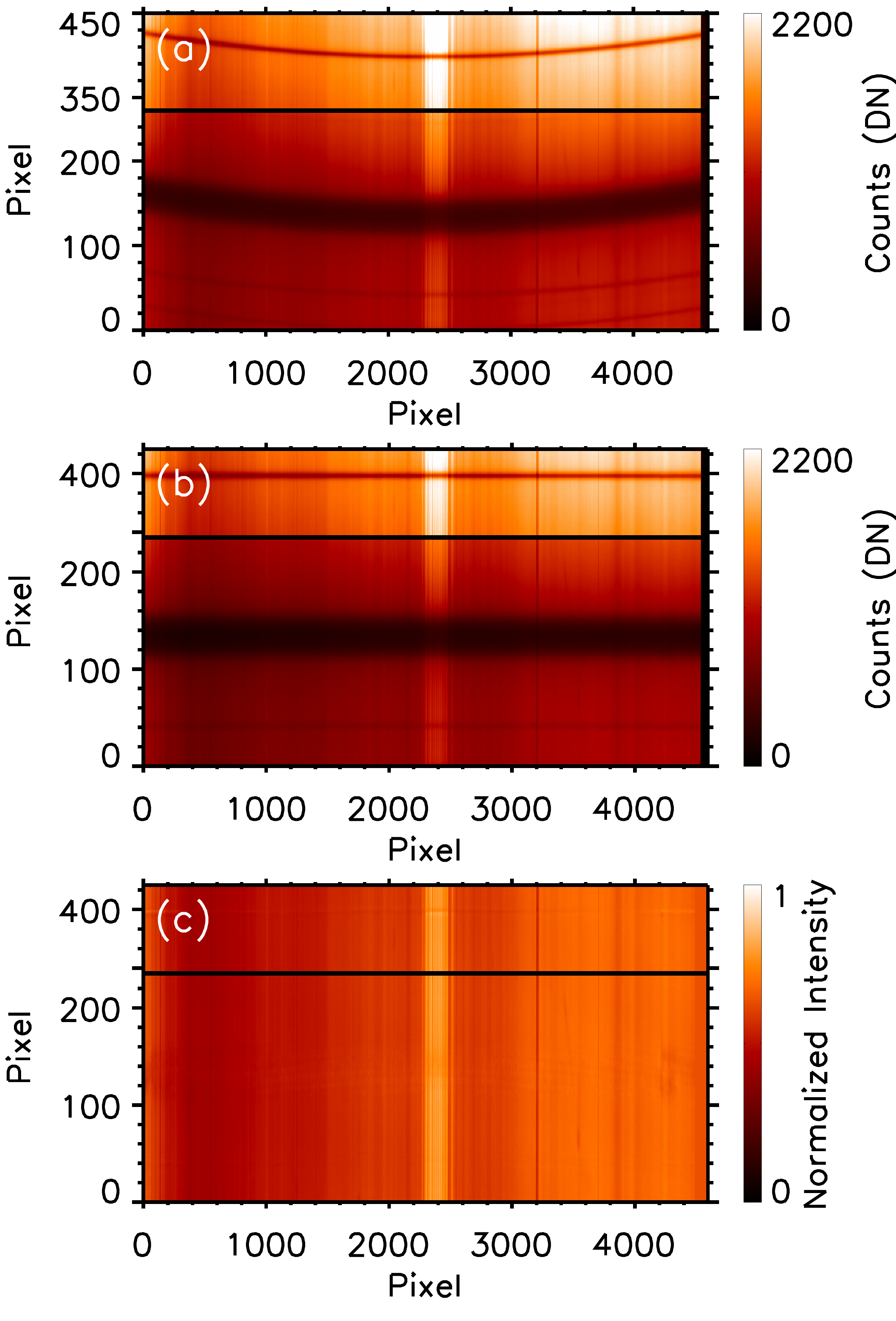}
	\caption{(a) The solar central spectrum stitched by a series of center-positioned spectra observed on December 23, 2021. (b) The solar central spectrum after correcting the slit image curvature. (c) The flat field of RSM.} 
	\label{fig:flat}
\end{figure}

\begin{figure*}[t]
	\centering
	\includegraphics[scale=0.55]{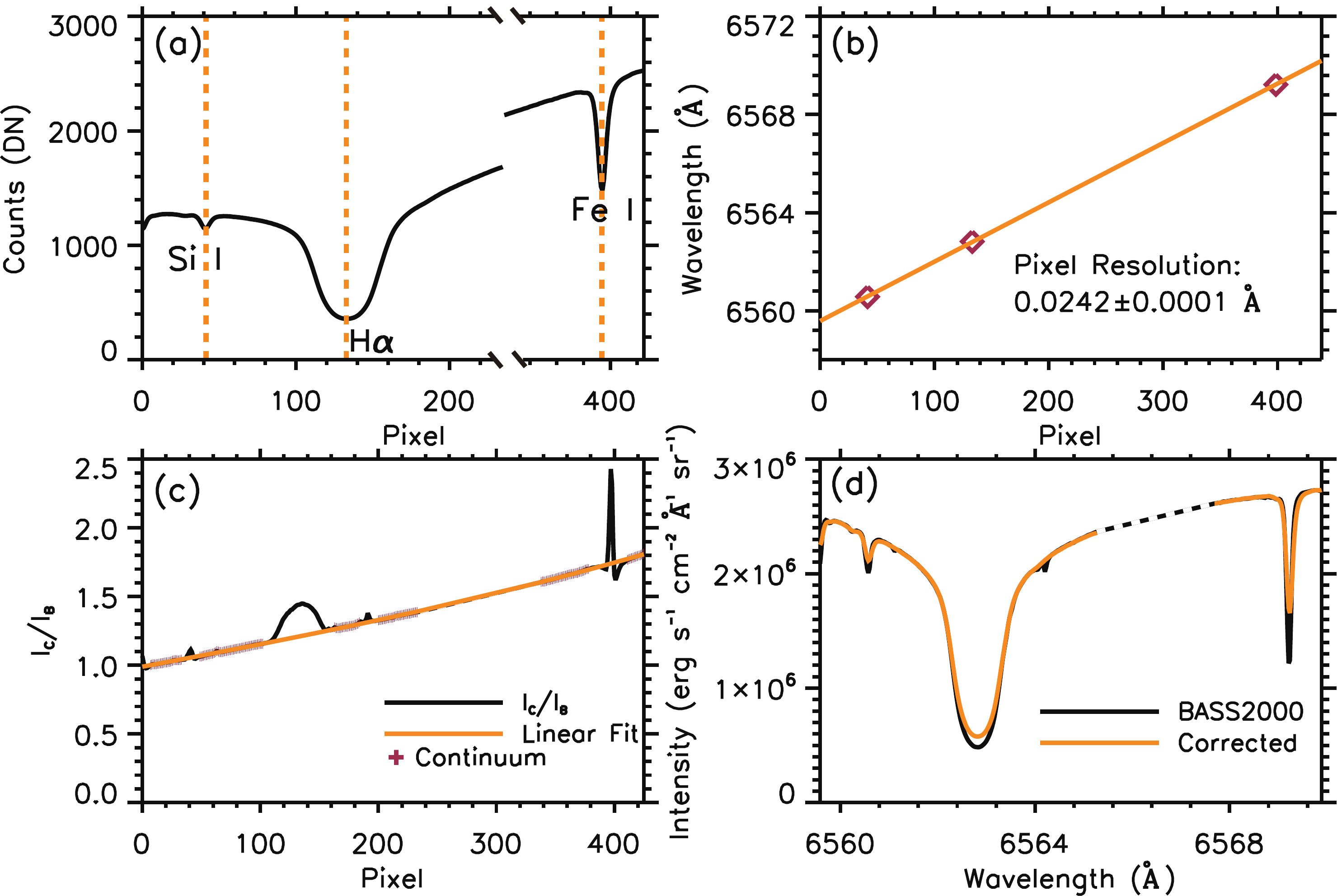}
	\caption{(a) Original spectral profiles of Si I, H$\alpha$ and Fe I lines. The dashed lines mark the positions of line cores. (b) Linear fitting of the wavelength with respect to the line core positions yielding the pixel spectral resolution. (c) Linear fitting of the continuum ratio between the CHASE/HIS spectra and the spectra of BASS2000, yielding a correction factor to the intensity non-uniformity along wavelength. (d) Comparison between the calibrated spectral profiles and the ones from BASS2000.} 
	\label{fig:calibration}
\end{figure*}

\subsection{Flat field and slit image curvature}

The flat field of RSM includes the systematic vignetting, the artifacts on the slit and detector, and the intensity patterns due to irregularities of the slit width. To obtain the flat field, we first gradually move the centeral part of the solar disk along the slit, during which a series of center-positioned spectra are recorded. This process was carried out in less than 1 minute by pointing the satellite from one side of the Sun along the diameter to the other side. The center-positioned spectra do not contain the information of solar limb darkening and solar differential rotation. They are then stitched into one solar central spectrum by using the weighted average fusion algorithm, as shown by Figure \ref{fig:flat}(a).

As shown in Figure \ref{fig:flat}(a), in the original slit-wavelength plot, the slit image is not straight but curved symmetrically. The slit image curvature arises from the effects of off-axis optics and the diffraction of plane grating \cite{zhao}, which is a common phenomenon in compact spectrographs with short focal lengths. The curvature of the slit image implies an offset of the observed wavelength with respect to the real one, which can be described by a quadratic function:
\begin{equation}\label{curve}
	\lambda' = c\lambda(x-x_0)^2 + \lambda,
\end{equation}
where $\lambda'$ and $\lambda$ are the observed and real wavelengths at specific pixel positions, $x_0$ is the pixel position of axis of symmetry in the bending slit image, and $c$ is a constant coefficient determined by the properties of the slit-raster system. It can be found that the larger the wavelength is, the stronger the bending of the slit image is. For each of the H$\alpha$, Si I and Fe I lines, we fit the curved slit image to get the position of the axis of symmetry. The constant coefficient $c$ is determined from the experimental data by using a wavelength tunable laser. After that, we can correct the curvature effect in the original slit-wavelength plot. Figure \ref{fig:flat}(b) shows the curvature-corrected solar central spectrum.

To derive the flat field, we need further to remove the spectral information of H$\alpha$, Si I and Fe I lines by using the so-called mean profile method \cite{bommier1,bommier2}. To do so, we divide the solar central spectrum by a mean spectrum that is an average over the slit. Figure \ref{fig:flat}(c) shows the final flat field of RSM based on the spectra of the central part of the Sun observed on December 23, 2021, which contains artificial features such as the bright and dark stripes due to irregularities of the slit, and the non-uniform responses on the detector. The dark and flat fields are scheduled to be updated once a month.

\subsection{Wavelength and intensities}

Figure \ref{fig:calibration}(a) shows the H$\alpha$, Si I and Fe I spectral profiles averaged near the solar disk center during the full-Sun scanning at 00:52:49 -- 00:53:35 UT on October 24, 2021. The dashed lines mark the line cores calculated by the centroid method. The pixel spectral resolution is derived to be $\sim$0.024 $\AA$ by performing a linear fitting to the air wavelengths of line cores as a function of the corresponding pixel positions, as shown in Figure \ref{fig:calibration}(b). It is consistent with the result obtained from experiments done in the laboratory. The instrument spectral resolution or the spectral FWHM is $\sim$0.072 $\AA$. The pixel spatial resolution is 0.52 arcsec.

In the observed spectra, there also appears a systematic increase in the intensity with wavelength, which is mainly caused by the instrument. To correct such an intensity non-uniformity, we assume a linear correction factor with wavelength. Figure \ref{fig:calibration}(c) shows the linear fitting to the ratio of continuum part of the CHASE/HIS spectra to that of the spectra of BASS2000$\footnote{https://bass2000.obspm.fr/solar\_spect.php}$. The fitting result can be used to correct the systematic intensity non-uniformity.

The corrected spectral profiles are shown as the orange curves in Figure \ref{fig:calibration}(d), which coincide well with the BASS2000 profiles, except for slightly wider wings and shallower cores. This small difference probably arises from the different instrumental profiles \cite{cai}. The absolute intensities of the spectra for the quiet Sun can be found in \textit{Allen's Astrophysical Quantities} \cite{cox}. For instance, the intensity at the continumm around 6500 $\AA$ is 2.88 $\rm \times 10^6 \ ergs^{-1} cm^{-2} $\AA$ \rm^{-1} sr^{-1}$. Therefore, the absolute intensities of the observed spectral lines of CHASE/HIS can be calibrated according to the continuum intensities.

\begin{figure}[H]
	\centering
	\includegraphics[scale=0.3]{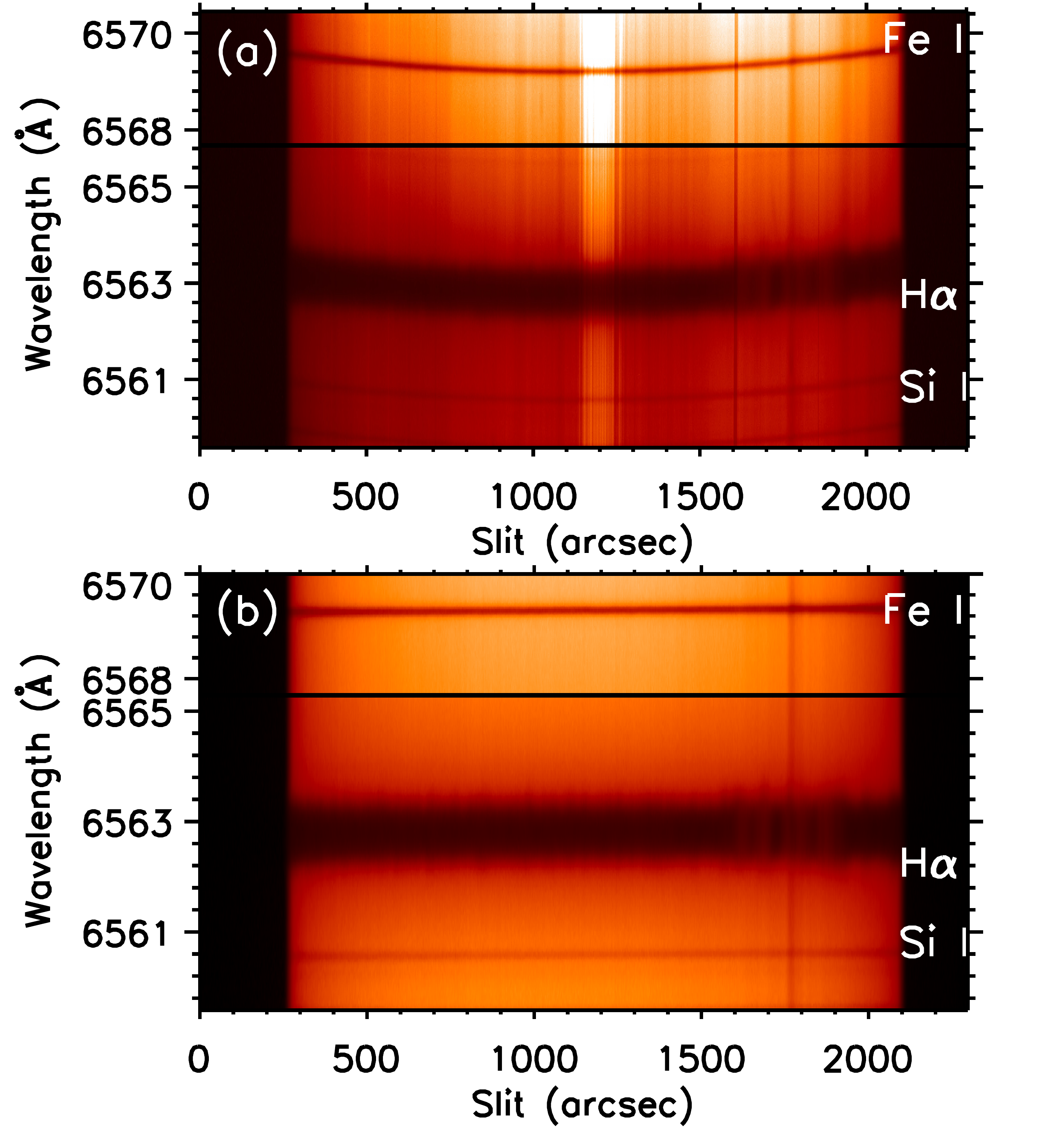}
	\caption{(a) The Level 0 RSM spectrum observed at 06:01:27 UT on December 22, 2021. (b) The calibrated Level 1 RSM spectrum.} 
	\label{fig:data}
\end{figure}

\subsection{Coordinates}

The quarternions are widely used as the attitude parameters of a satellite or a payload. They are defined as $q_{0}=\cos(\mu/2)$, $q_{1}=l\sin(\mu/2)$, $q_{2}=m\sin(\mu/2)$, and $q_{3}=n\sin(\mu/2)$, where $\mu$ represents the rotation angle of the vector ($l,m,n$). Using the quarternions of CHASE/HIS, we can derive the transformation from the instrumental coordinates to the J2000 equatorial coordinates \cite{aoki}. To do this, we first obtain the three attitude angles of the CHASE/HIS by using the expressions of $\gamma = - \arcsin (2q_{1}q_{3} - 2q_{0}q_{2})$, $\theta = \arctan [(2q_{2}q_{3} + 2q_{0}q_{1}) / (q_{0}^{2} - q_{1}^{2} - q_{2}^{2} + q_{3}^{2})]$, and $\psi = \arctan [(2q_{1}q_{2} + 2q_{0}q_{3}) / (q_{0}^{2} + q_{1}^{2} - q_{2}^{2} - q_{3}^{2})]$. The instrumental coordinates are then converted to the J2000 equatorial coordinates via the rotation matrix

\footnotesize
\begin{equation}\label{matrix}
	  R_{rot} = \begin{pmatrix}
			\cos \gamma \cos\psi & \sin\theta \sin\gamma \cos\psi - \cos\theta \sin\psi  & \cos\theta \sin\gamma \cos\psi + \sin\theta \sin\psi \\
			\cos\gamma \sin\psi & \sin\theta \sin\gamma \sin\psi + \cos\theta \cos\psi & \cos\theta \sin\gamma \sin\psi - \sin\theta \cos\psi \\
			- \sin\gamma & \sin\theta \cos\gamma & \cos\theta \cos\gamma
		\end{pmatrix}.
\end{equation}
\normalsize
If a normalized vector on the image observed by CHASE/HIS is expressed as ($x$, $z$)$^{T}$, then
\begin{equation}
	\begin{pmatrix}
	cos\delta cos\alpha \\ cos\delta sin\alpha\\ sin\delta
	\end{pmatrix}
    = R_{rot}
	\begin{pmatrix}
	x \\ 0 \\ z
    \end{pmatrix},
\end{equation}
where $\delta$ and $\alpha$ refer to the latitude and longitude of this vector in the J2000 equatorial coordinates. It should be notice that, if all the three attitude angles are set to be zero degree, the instrument pointing is parallel to the y-axis of the J2000 coordinates with the top of images on the detector parallel to the z-axis.

After transforming the CHASE/HIS coordinates, we can determine the plane of sky of the detector, through which we calculate the B0 angle, i.e., the heliographic latitude of the solar image center, and the image rotation angle relative to the Solar North Pole (SNP). Here we adopt the SNP ($\alpha_{0}$ = 286.13$^\circ$, $\delta_{0}$ = 63.87$^\circ$) given by Archinal et al. \cite{archinal}. In the FITS file headers of the Level 1 data, the key word $\textbf{INST\_ROT}$ refers to the clockwise angle of the SNP relative to the image top of the CHASE/HIS, and the keyword $\textbf{B0}$ refers to the heliographic latitude of the solar disk center.

\subsection{RSM data products}

Figure \ref{fig:data}(a) shows an example of the Level 0 RSM spectrum observed at 06:01:27 UT on December 22, 2021, when the slit was scanning over the solar disk center. The Level 1 RSM spectrum is produced from the Level 0 data via the calibration procedures introduced above, including dark-field and flat-field correction, slit image curvature correction, wavelength and intensity calibration, and coordinate transformation. Figure \ref{fig:data}(b) shows the Level 1 RSM spectrum, from which one can clearly find the features of the solar limb darkening, an active region and a sunspot. The present calibration procedures do not include the correction of bad pixels or spikes arising from high-energy particles. It should be considered in near future according to the actual situations of particle contamination.

\begin{figure*}[t]
	\centering
	\includegraphics[scale=0.4]{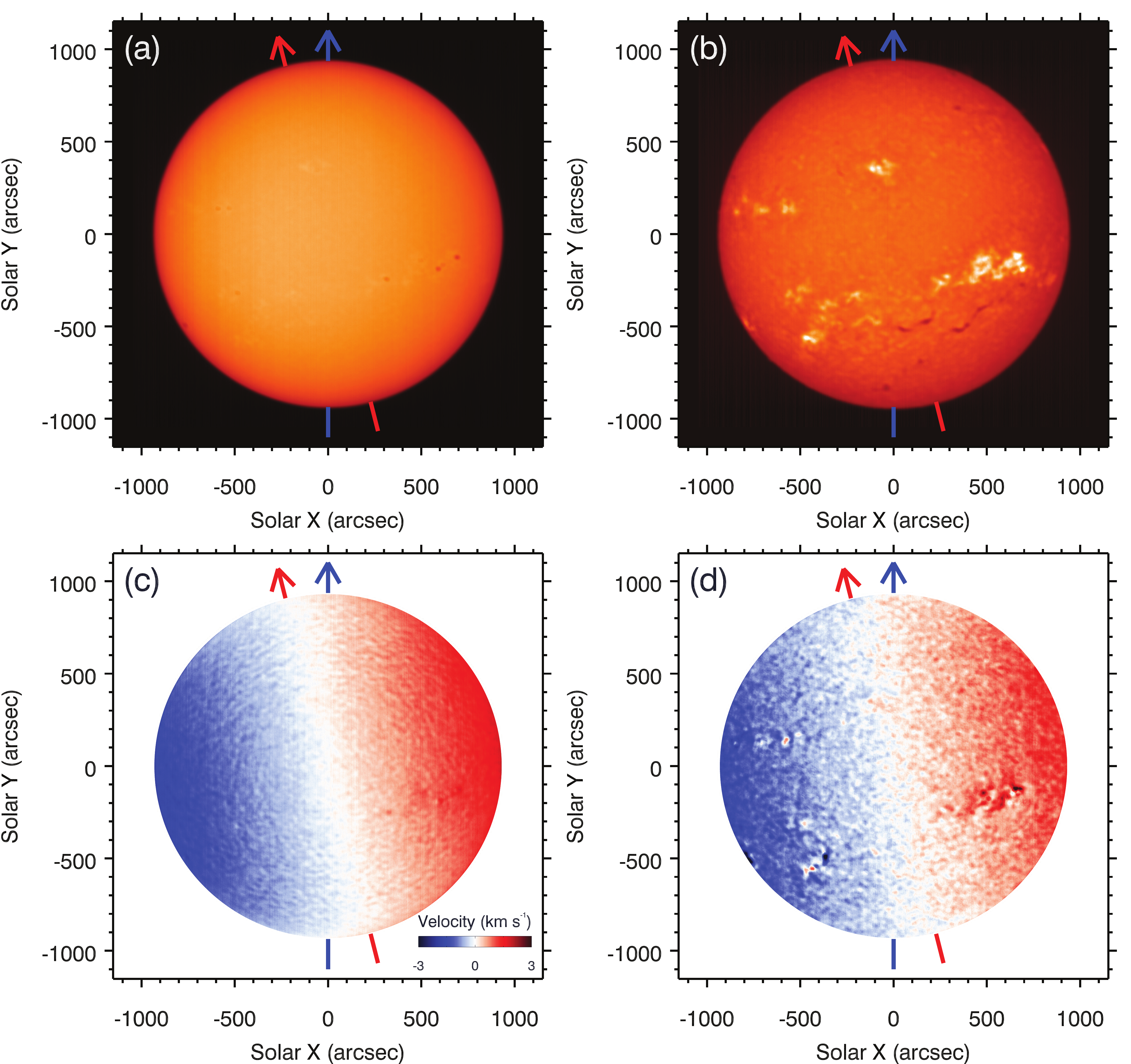}
	\caption{(a) The full-Sun photospheric image at 6569.2 $\AA$ on 2021 December 22. (b) The chromospheric image at 6562.8 $\AA$. (c) The full-Sun photospheric Dopplergram derived by using the Fe I spectral line. (d) The full-Sun chromospheric Dopplergram derived by using the H$\alpha$ spectral line. The blue and red arrows indicate the y-axis of the detector and the solar north pole. The angle between these two arrows is 14.1$^\circ$.} 
	\label{fig:full_disk}
\end{figure*}

The Level 1 RSM spectra are separately archived as two 3-dimensional FITS files for the H$\alpha$ and Fe I passbands, respectively. The three dimensions refer to the pixels along the slit, the wavelength, and the scanning steps, respectively. For a specific wavelength, one can plot a full-Sun or region-of-interest image at this wavelength. Figure \ref{fig:full_disk}(a) and \ref{fig:full_disk}(b) display, for example, a photopheric image at 6569.2 $\AA$ and a chromospheric image at 6562.8 $\AA$ observed at 06:01:05 -- 06:01:52 UT on November 22, 2021. One can find clearly the sunspots, filaments, and plages. The supplementary Movies 1 and 2 display the full-Sun spectroscopic images in the wavebands of H$\alpha$ and Fe I, respectively.

Based on the Level 1 RSM spectra, higher-level products can be further derived for spectific scientific objectives. For example, we can derive simultaneously the full-Sun or region-of-interest Dopplergrams in the photosphere and the chromosphere. The high spectral resolution of HIS instrument can sure a high accuracy in the derived velocity field. Figure \ref{fig:full_disk}(c) and \ref{fig:full_disk}(d) display the full-Sun photospheric and chromospheric Dopplergrams. Note that the Doppler velocities are calculated by using the cross-correlation method. The reference profile is taken to be an average profile over a central region of the solar disk. The accuracy of the velocity field is estimated to be $\sim$0.06 km s$^{-1}$. Therefore, the photospheric and chromospheric differential rotation from the poles to the equator can be clearly distinguished.

\section{Processing pipeline for CIM data}\label{sect4}

The CIM Level 0 data are archieved as 5120 $\times$ 5120 arrays. The temporal resolution of the CIM is 1 second, and the pixel spatial resolution is 0.52 arcsec. After corrections for the dark field, flat field and transformation to heliographic coordinates, the CIM Level 1 data are released to the users.

\begin{figure*}[t]
	\centering
	\includegraphics[scale=0.4]{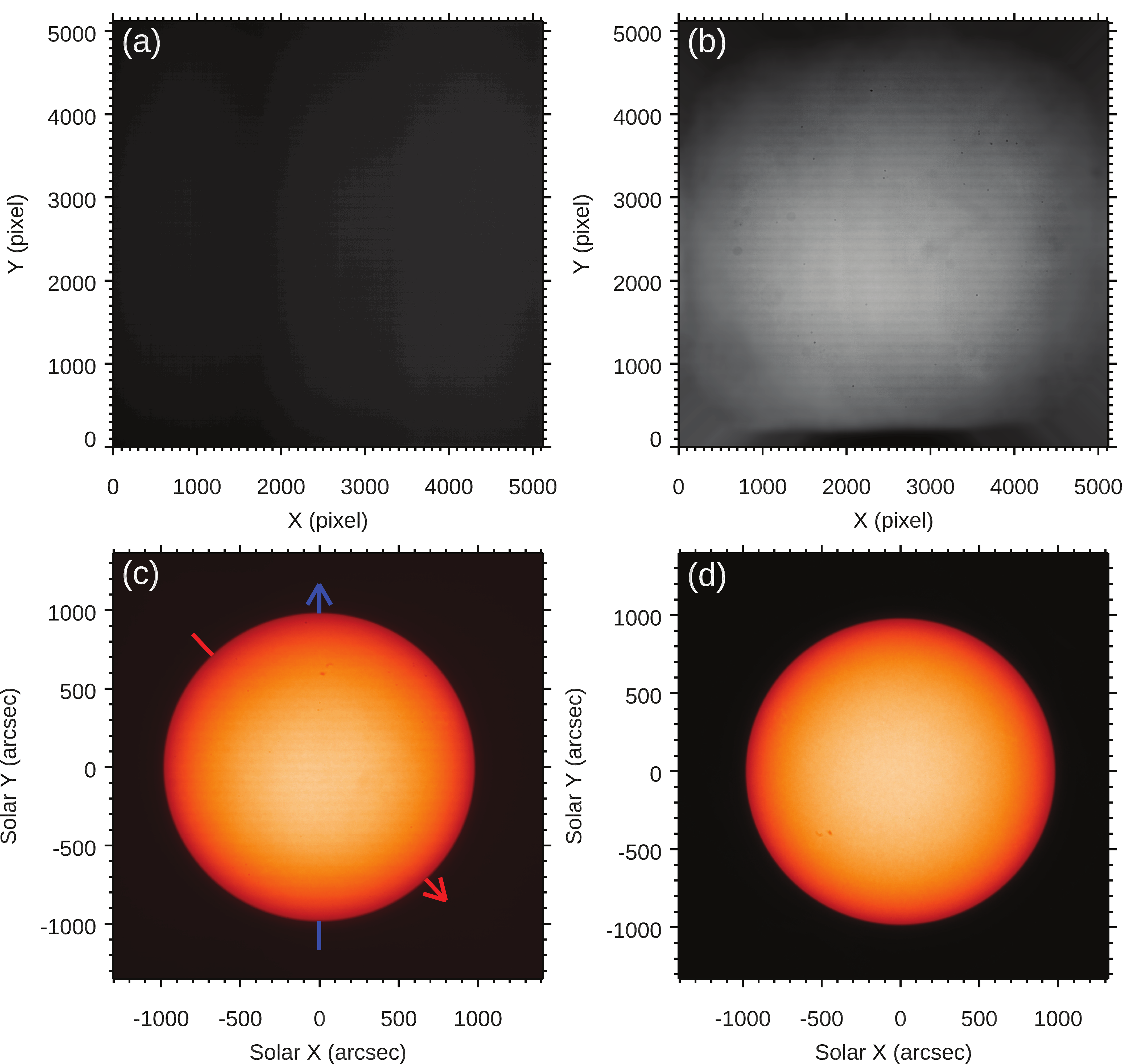}
	\caption{(a) The dark field of CIM obtained on 24 November 2021. (b) The flat field of CIM. (c) Original photospheric image observed at 02:02:40 UT on November 25, 2021. (d) Calibrated photospheric image. The blue and red arrows indicate the y-axis of the detector and the solar north pole. The angle between these two arrows is 136.7$^\circ$.} 
	\label{fig:CIM}
\end{figure*}

The dark field of CIM is obtained by using the same method as the RSM. The flat field of CIM is calculated with the KLL method \cite{kll}, which was widely applied in several other solar space missions \cite{sujt,feng,solanki,ljw}. This method requires multiple images centered at different positions of the Sun by changing the satellite pointing. Here we take nine images with one positioned at the disk center and the other eight at the limb. All the images are taken within 15 minutes, during which we assume no activities occur in the photosphere. These images are then input into the KLL iterative algorithm to produce the flat-field image. We carry out 10 iterations for each flat-field image. Figure \ref{fig:CIM}(a) and \ref{fig:CIM}(b) show the CIM dark field and flat field obtained on November 14, 2021. The CIM Level 0 image and calibrated Level 1 image are displayed in Figure \ref{fig:CIM}(c) and \ref{fig:CIM}(d), respectively. The dark-field and flat-field images are scheduled to be updated once a month.

\section{Summary}\label{summary}

The CHASE satellite was successfully launched on October 14, 2021. Its first-light imaging was obtained on October 24, 2021. Since then, the HIS intrument onboard the CHASE mission operates well as expected. After a commission phase of about 5 months, CHASE has started its routine observations. The first set of data has been calibrated and analyzed recently. The Level 1 science data in FITS format will be available to the community through the website at SSDC-NJU (\url{https://ssdc.nju.edu.cn}).

\Acknowledgements{The CHASE mission was supported by the China National Space Administration (CNSA).}




\begin{thebibliography}{99}

\bibitem{hale}  G. E. Hale, 1908, Contributions from the Mount Wilson Observatory, \textit{Astrophys. J.} \textbf{27}, 219

\bibitem{chen1} P. F. Chen, A. A. Xu, and M. D. Ding, 2020, \textit{Research in Astron. Astronphys.} \textbf{20}, 166

\bibitem{ichimoto} K. Ichimoto and H. Kurokawa, 1984, \textit{Solar Phys.} \textbf{93}, 105

\bibitem{Berlicki} A. Berlicki, 2007, \textit{The Physics of Chromospheric Plasmas, Astronomical Society of the Pacific Conference Series}, \textbf{368}, 387

\bibitem{Madjarska} M. S. Madjarska, J. Chae, F. Moreno-Insertis, Z. Hou, D. Nobrega-Siverio, H. Kwak, K. Galsgaard, and K. Cho, 2021, \textit{Astron. Astrophys.} \textbf{646}, A107

\bibitem{hong} J. Hong, Y. Li, M. D. Ding, and Q. Hao, 2021, \textit{Astrophys. J.}, \textbf{921}, 50

\bibitem{fang1} C. Fang, P. F. Chen, Z. Li, M. D. Ding, Y. Dai, X. Y. Zhang, et al., 2013, \textit{Research in Astron. Astronphys.} \textbf{13}, 1509

\bibitem{liu} Z. Liu, J. Xu, B. Z. Gu, S. Wang, J. Q. You, L. X. Shen, et al., 2014, \textit{Research in Astron. Astronphys.} \textbf{14}, 705

\bibitem{potzi} W. P\"{o}tzi, A. M. Veronig, M. Temmer, D. J. Baumgartner, H. Freislich, and H. Strutzmann, 2016, \textit{Solar Phys.} \textbf{291}, 3103

\bibitem{fang2}  C. Fang, B. Gu, X. Yuan, M. Ding, P. Chen, Z. Dai, et al., 2019, \textit{Sci. China-Phys. Mech. Astron.} \textbf{49}, 059603

\bibitem{zhangp} P. Zhang, X. Hu, Q. F. Lu, A. J. Zhu, M. Y. Lin, L. Sun, et al., 2022, \textit{Adv. Atoms. Sci.} \textbf{39}, 1

\bibitem{li1} C. Li, C. Fang, Z. Li, M. D. Ding, P. F. Chen, Z. Chen, et al., 2019, \textit{Research in Astron. Astronphys.} \textbf{19}, 165

\bibitem{li2} C. Li, C. Fang, Z. Li, M. D. Ding, P. F. Chen, Y. Qiu, et al., 2022, \textit{Sci. China-Phys. Mech. Astron.}, \textbf{65}, 289602

\bibitem{gan} W. Q. Gan, C. Zhu, Y. Y. Deng, H. Li, Y. Su, H. Y. Zhang, et al., 2019, \textit{Research in Astron. Astronphys.} \textbf{19}, 156

\bibitem{zhang1} W. Zhang, W. Cheng, W. You, X. Chen, J. Zhang, C. Li, and C. Fang, 2022, \textit{Sci. China-Phys. Mech. Astron.}, \textbf{65}, 289604

\bibitem{han} Q. Liu, H. J. Tao, C. Z. Chen, C. S. Han, Z. Chen, and G. Mei, 2022, \textit{Sci. China-Phys. Mech. Astron.}, \textbf{65}, 289605

\bibitem{zhao} J. Zhao, 2003, \textit{Applied Spectroscopy}, \textbf{57}, 1368

\bibitem{bommier1} V. Bommier and G. Molodij, 2002, \textit{Astron. Astrophys.} \textbf{381}, 241

\bibitem{bommier2} V. Bommier and J. Rayrole, 2002, \textit{Astron. Astrophys.}, \textbf{381}, 227

\bibitem{cai} Y. Cai, Z. Xu, and K. Ji, 2020, \textit{Solar Phys.} \textbf{295}, 31

\bibitem{cox} A. N. Cox, 1999, \textit{Allen's Astrophysical Quantities}, Springer-Verlag, New York, Inc.

\bibitem{aoki} S. Aoki, M. Soma, H. Kinoshita, K. Inoue, 1983,  \textit{Astron. Astrophys.} \textbf{128}, 263

\bibitem{archinal} B. A. Archinal, C. H. Acton, M. F. AHearn, A. Conrad, G. J. Consolmagno, T. Duxbury, et al., 2018, \textit{Celestial Mech. Dyn. Astron.} \textbf{130}, 22

\bibitem{kll} J. R. Kuhn, H. Lin, and D. Loranz, 1991, \textit{PASP}, \textbf{103}, 1097

\bibitem{sujt} J. T. Su, X. Y. Bai, J. Chen, J. J. Guo, S. Liu, X. F. Wang, et al., 2019, \textit{Research in Astron. Astronphys.}, \textbf{19}, 161

\bibitem{feng} L. Feng, H. Li, B. Chen, Y. Li, R. Susino, Y. Huang, et al., 2019, \textit{Research in Astron. Astronphys.}, \textbf{19}, 162

\bibitem{solanki} S. K. Solanki, J. C. del Toro Iniesta, J. Woch, A. Gandorfer, J. Hirzberger, A. Alvarez-Herrero, et al., 2020, \textit{Astron. Astronphys.}, \textbf{642}, A11

\bibitem{ljw} J. W. Li, H. Li, L. Feng, Y. Huang, J. Zhao, L. Lu, et al., 2021, \textit{Research in Astron. Astronphys.}, \textbf{21}, 121
	
\end{thebibliography}

\end{multicols}
\end{document}